\documentstyle[12pt,amstex]{article}
\newcommand{\RR}{{\Bbb R}}                      
\newcommand{\CC}{{\Bbb C}}                      

\newcommand{\la}{{\lambda}}                     
\newcommand{\Oo}{{\cal O}}                      
\newcommand{\AO}{{\cal A}({\cal O})}            

\newcommand{\Al}{{\cal A}}                      
\newcommand{\All}{{\cal A}^{(\lambda)}}         
\newcommand{\AlO}{{\cal A}^{(\lambda)} ({\cal O})} 
\newcommand{\Alu}{\underline{{{\cal A}_{ }}}}         
\newcommand{\AOu}{\underline{{{\cal A}_{ }}} ({\cal O})}

\newcommand{\Au}{{\underline{{A}_{ }}}}                 
\newcommand{\Aul}{{\underline{{A}_{ }}}_{\lambda}}    
\newcommand{\Pg}{{\cal P}_{+}^{\uparrow}}       
\newcommand{\Lx}{( \Lambda , x )}               
\newcommand{\aLx}{{\alpha}_{\Lambda , x}}       
\newcommand{\alLx}{{\alpha}^{(\la)}_{\Lambda , x}}       
\newcommand{\al}{{\alpha}^{( \lambda )}}        
\newcommand{\auLx}{{\underline{{\alpha}_{ }}}_{\Lambda, x}}
\newcommand{\Ho}{{\cal H}_{\omega}}             
\newcommand{\Ooo}{{{\Omega}_{\omega}}}          
\newcommand{\po}{{\pi}_{\omega}}                

\newcommand{\ou}{{\underline{\omega_{ }}}}          
\newcommand{\pu}{{\underline{\pi}}}             

\begin{document}
\title{Short Distance Analysis\\
 in Algebraic Quantum Field Theory\thanks{Invited talk at International
Congress of Mathematical Physics, July 1997, Brisbane}}
\author{Detlev Buchholz
\\[4mm]
Institut f\"ur Theoretische Physik, Universit\"at 
G\"ottingen,\\ Bunsenstra{\ss}e 9, D-37073 G\"ottingen, Germany\\[2mm]} 
\date{}
\maketitle
\abstract{\noindent Within the framework of algebraic quantum field theory 
a general method is presented which allows one to compute and
classify the short distance (scaling) limit of any algebra of 
local observables. The results can be used to determine the particle 
and symmetry content of a theory at very small scales and 
thereby give an intrinsic meaning to notions such as ``parton'' 
and ``confinement''. The method has been tested in models.}
\section{Introduction}
Within the Lagrangean approach to quantum field theory, a powerful tool
for the analysis of the ultraviolet properties of models is based
on scaling (renormalization group) transformations. They
allow one to interpret the theory at small spacetime scales and have led to 
fundamental concepts such as the notion of parton, confinement, 
asymptotic freedom etc. 

This approach has been substantial for the present theoretical
understanding of high energy physics. Nevertheless, it is not completely
satisfactory since it is based on the (gauge) fields appearing in the  
Lagrangean which in general do not admit a direct physical
interpretation. It therefore seems desirable to establish a framework 
for the short distance analysis
which relies on observables only. Since observables are typically
composed of several elementary fields this may not seem an easy
task if one thinks e.g.\ of a generalization of renormalization group
equations. But it turns out that there is a conceptually simple
solution of this problem \cite{BuVe1}
in the general algebraic framework of local quantum 
physics \cite{Ha}. 

In \cite{BuVe1} we have established a method which allows
one to define scaling transformations and the scaling limit for any
given algebra of local observables $\Al$. For the convenience of the
reader who is not familiar with the algebraic setting, 
we briefly list the relevant assumptions. 

1.  {(Locality)} We suppose
that the local observables of the underlying theory generate a net
of local algebras over $d$-dimensional Minkowski space $\RR^{\, d}$, i.e.\ an
inclusion preserving map
$$ \Oo \rightarrow \AO $$
from the set of open, bounded regions
$\Oo$
in Minkowski space to unital C$^*$-algebras
$\AO$.
The algebra generated by all local algebras
$\AO$
(as a C$^*$-inductive limit) will be denoted by
$\Al$.
The net is supposed to satisfy the principle of locality (Einstein
causality), i.e.\ operators localized in spacelike separated regions
commute. 

2. {(Covariance)} The Poincar\'e group
$\Pg$
is represented by automorphisms of the net. Thus for each
$\Lx \in \Pg$
there is an
$\aLx \in \mbox{Aut} \Al$
such that, in an obvious notation,
$$ \aLx ( \AO ) = \Al ( \Lambda \Oo + x ) $$
for any region
$\Oo$.
We amend this fundamental postulate by a continuity condition 
and assume that for any
$A \in \Al$
the function
$\Lx \rightarrow \aLx (A)$
is strongly continuous. 

3. {(States)} Physical states are described by positive, linear
and normalized functionals
$\omega$
on
$\Al$.
By the GNS-construction, any state
$\omega$
gives rise to a representation
$\po$
of
$\Al$
on a Hilbert space
$\Ho$,
and there exists a vector
$\Ooo \in \Ho$
such that
$$ \omega (A) \ = \ ( \Ooo, \po (A) \Ooo ), \ A \in \Al. $$ 
The state describing the vacuum is distinguished by the fact 
that, on the corresponding Hilbert space $\Ho$, 
there is a continuous unitary representation $U_\omega \Lx $
of the Poincar\'e group $\Pg$ 
which leaves the vector $\Ooo $ invariant, satisfies the
relativistic spectrum condition (positivity of energy) 
and implements the action of $\Pg$ on the
observables, 
$$ U_\omega \Lx   \po (A) { U_\omega  \Lx  }^{-1} \ = \ \po ( \aLx
(A)),  \, A  \in \Al. $$ 
Any state of physical interest is assumed to be locally normal 
to the vacuum state.

\section{Scaling algebra and scaling limit}
What is required in order to carry over the ideas of renormalization 
group analysis to this algebraic setting? One first has to proceed from the 
given net and automorphisms $\Al, \alpha$ at spacetime scale $\la
= 1$ (in appropriate units) 
to the corresponding nets $\All, \al$ describing the theory at
arbitrary scale $\la \in \RR_+$. This is easily accomplished by
setting for given $\la$
$$ \AlO \doteq \Al ( \la \Oo ), \quad \alLx \doteq \alpha_{\Lambda,
  \la x}.$$
Note that the latter nets are in general not isomorphic to the
original one. They are to be regarded as different theories with scaled mass
spectrum, running coupling constants etc. 

In addition to this passage from the given theory to the 
corresponding theories at arbitrary scale 
one needs a way of comparing the respective observables. This 
is accomplished by considering functions $\Au$
of the scaling parameter with values in the algebra of observables $\Al$. 
For given $\la$ the value 
$\Au_\la$ of $\Au$ is to be regarded as an observable in the
theory $\All, \al$. The graph of the function $\Au$ thus  
provides the desired identification of observables at different
scales. With this idea in mind one is led to the concept of scaling 
algebra.

The {\em scaling algebra\/} 
$\Alu$
associated with any given local net
$\Al$ consists of functions $ \Au : \, \RR_+ \rightarrow \Al$.
The algebraic operations in
$\Alu$
are pointwise defined by the corresponding operations in
$\Al$,
and there is a C$^*$-norm on
$\Alu$
given by
$$ || \Au || = \sup_{\la} || \Aul ||. $$
The local structure of
$\Al$
is lifted to
$\Alu$
by setting
$$ \AOu = \{\Alu : \, \Aul \in \AlO , \, \la \in \RR_+
\}. $$
Hence
$\Oo \rightarrow \AOu $
defines a net over Minkowski space and 
$\Alu$ is defined as the C$^*$-inductive limit of the local algebras
$\AOu$.
It is easily verified that this net is local.
Moreover, the action of the Poincar\'e group in the underlying theory 
can be lifted to an action of automorphisms 
$\auLx$ on 
$\Alu$ which is given by
$$ \big( \auLx ( \Au ) \big)_\la \doteq \alLx ( \Aul ). $$
It is assumed that $\Alu$ consists only of elements on which 
these automorphisms act strongly continuously, i.e.\ 
$$
||\,  \auLx ( \Au ) - \Au \, || \rightarrow 0 \quad \mbox{as} \quad \Lx 
\rightarrow (1,0).
$$
Heuristically speaking, the latter constraint amounts to
the condition \cite{BuVe1} 
that the operators $\Aul$ in the graph of a given 
element $\Au \in \Alu$ occupy, for 
all values of $\la$, a fixed volume of ``phase space''. Hence, 
whereas the scale of spacetime changes 
along the graph, the scale $\hbar$  of action
is kept fixed. 

The structure of the physical states $\omega$ in the underlying theory
at small spacetime scales can now be analyzed with the help of the
scaling algebra as follows. Given $\omega$, one defines the lift of 
this state to the scaling algebra at scale $\la \in \RR_+$ by setting 
$$ \ou_\la ( \Au ) \doteq \omega ( {\Au}_{\lambda} ), \quad \Au \in
\Alu. $$
Let $\pu_\la$ be the GNS--representation of $\Alu$ which is induced 
by $\ou_\la$. We then consider the net 
$${\cal O} \rightarrow \underline{{\cal A}_{ }} ({\cal O})/ \mbox{ker}
\underline{\pi_{ }}_\lambda, \quad \underline{\alpha_{ }}^{(\lambda)}, $$
where $\mbox{ker}$ means ``kernel'' and  $\underline{\alpha_{ }}^{(\lambda)} $
is the induced action of the Poincar\'e transformations 
$\underline{\alpha_{ }}$ on this quotient. It is important to note 
\cite{BuVe1} that this net is isomorphic to the underlying theory 
$\All, \al$ at scale
$\la$. This insight leads to the following canonical definition of the
scaling limit of the theory: One first considers the limit(s) of the
net of states
$\{ \underline{\omega_{ }}_\lambda \}_{\lambda \searrow
    0}$. By standard compactness arguments, this net 
has always a non--empty
set $ \{ \underline{\omega_{ }}_0 \}  $ of limit points. The following
facts about these limit states have been established in \cite{BuVe1}:\\[2mm] 
{\small 1.} \, The set $ \{ \underline{\omega_{ }}_0 \}  $ does not
depend on the chosen physical state $\omega$.\\[2mm]
{\small 2.} \, Each $ \underline{\omega_{ }}_0$ is a vacuum
state on $\Alu$ which is pure in $d>2$ spacetime dimensions.\\[2mm]
Denoting the GNS--representation corresponding to given 
$\underline{\omega_{ }}_0 $ by $(\underline{\pi_{ }}_0, \, 
  \underline{{\cal H}_{ }}_0)$ one can then define in complete
analogy to the case $ \la > 0$ the net 
$$ {\cal O} \rightarrow 
{\cal A}^{(0)} \doteq  \underline{{\cal A}_{ }} ({\cal O})/
\mbox{ker} \, \underline{\pi_{ }}_0, \quad 
\alpha^{(0)} \doteq \underline{\alpha_{ }}^{(0)} $$ 
and the corresponding vacuum state $\omega_0 \doteq \mbox{proj} \, 
\underline{\omega_{ }}_0$, where $\mbox{proj}$ denotes the projection
of the respective state to the quotient algebra. This net is
to be interpreted as {\em scaling limit of the underlying theory}.  
It should be remarked that in view of the possible appearance of several 
limit points $ \underline{\omega_{ }}_0$ this limit may not be
unique. 

The preceding steps which have led us from the original net 
to its scaling limit(s) are pictorially described in 
the diagram 
$$ {\cal A}, \, \alpha \longrightarrow \underline{{\cal A}_{ }}, \,
\underline{\alpha_{ }} \longrightarrow \{ {\cal A}^{(0)}, \, \alpha^{(0)}
\}. $$

\section{Classification of theories}
The scaling limits of local nets of observable algebras can be
classified according to the following three mutually exclusive 
general possibilities.\\[2mm]
{\bf \noindent Classification:} Let $\Al, \alpha$ be a net 
with properties
specified in the Introduction. There are the following possibilities for
the structure of the corresponding scaling limit theory.
\begin{itemize}
\item[{\small 1.}] The nets $\{ {\cal A}^{(0)}, \, \alpha^{(0)} \}$
are isomorphic to the trivial net $ \{ {\CC} \! \cdot \! 1, \mbox{id} \} $
(classical scaling limit) 
\item[{\small 2.}] The nets $\{ {\cal A}^{(0)}, \, \alpha^{(0)} \}$ are
  isomorphic and non-trivial (quantum scaling limit)
\item[{\small 3.}] Not all of the nets $\{ {\cal A}^{(0)}, \, \alpha^{(0)} \}$
are isomorphic (degenerate scaling limit)
\end{itemize}
\vspace*{2mm}
In theories with a classical scaling limit 
all correlations between quantum observables in $\Alu$ disappear at
small scales, hence the terminology. 
Recently, examples of local nets with a classical scaling limit have 
been constructed in \cite{Lu}. They satisfy the standard
conditions for nets of physical interest, such as weak additivity,
wedge-duality, nuclearity etc. But in contrast to nets generated by 
Wightman fields, they contain only operators which exhibit a very singular
(non--temperate) short distance behavior. To some extent these
examples mimic the
ultraviolet problems which one expects to meet in non--renormalizable 
theories or theories without ultraviolet fixed point. It is of
interest that theories with a classical scaling limit 
can be characterized by their phase space properties 
in terms of nuclearity criteria \cite{Bu1}. 

The case of theories with a quantum scaling limit is expected to be  
the generic one. Here the terminology derives from the fact that the
algebras ${\cal A}^{(0)} $ are necessarily non--abelian if they are 
non--trivial \cite{Bu1}. Simple examples in this class are free field 
theories in $d = 3$ and $4$ dimensions 
as well as some exactly solvable models for $d = 2$ \cite{BuVe2}. 
It seems likely that 
quantum field theories with a stable ultraviolet fixed point, e.g.\ 
asymptotically free theories, also belong to this class. A
clarification of this point would be very desirable. 

In theories with a degenerate scaling limit it is not possible to
describe the short distance properties in terms of a single scaling
limit theory, the structure continually varies as $\la$ approaches
$0$. Candidates for this type of behavior are 
theories with a large number of local degrees of freedom which
strongly violate the above mentioned nuclearity criteria. 
Yet these results are not yet complete. 
It would be of great interest, both from a conceptual and technical
point of view,  
to clarify the relation between the phase space properties of the underlying
theory and the nature of its scaling limit.

Since theories with a quantum scaling limit are of particular
interest let us mention the following two general facts which
have been established in \cite{BuVe1}:
\begin{itemize}
\item[{\small 1.}] In any theory with a quantum scaling limit the
limit net
${\cal A}^{(0)}, \, \alpha^{(0)}$ is dilatation invariant.
\item[{\small 2.}] If this limit net satisfies the Haag--Swieca
compactness criterion (which is the case whenever the underlying
theory complies with a quantitative version of this criterion \cite{Bu1})
then the scaling limit of ${\cal A}^{(0)}, \, \alpha^{(0)}$   is 
isomorphic to itself, i.e.\ it is a fixed point under the above procedure.
\end{itemize}

\section{Ultraparticles and ultrasymmetries}
The fact that the scaling limit theories ${\cal A}^{(0)}, \,
\alpha^{(0)}$ have all features 
of a local net of observable algebras allows one to introduce standard 
concepts for their physical interpretation. We call the particles 
appearing in the scaling limit theory (in the sense of Wigner's particle 
concept) {\em ultraparticles} and the global gauge symmetries of the 
scaling limit theory {\em ultrasymmetries}. 

In order to determine the particle and symmetry content in the scaling
limit from the 
net ${\cal A}^{(0)}, \, \alpha^{(0)}$ 
one has to apply the Doplicher--Roberts reconstruction theorem. By 
this method one can recover    
the physical Hilbert space of all states carrying a localizable 
charge, the algebra of charge carrying fields and the global gauge
group \cite{DoRo}. The necessary prerequisite for this construction, 
Haag--duality, is given in the scaling limit whenever  
the underlying theory complies with the
special condition of duality invented by Bisognano and Wichmann 
\cite{BuVe1}. 

Of particular interest is the comparison of the particle and symmetry
content of the underlying theory and of its scaling limit. There may
be particles which disappear in the scaling limit (think of ``hadrons''), 
particles which survive (``leptons'') and particles which
only come into existence at small scales (``quarks'', ``gluons'').\\
{\unitlength1cm
\begin{picture}(10,4.5)
\thicklines 

\put(7.5,3){\oval(6,1.5)} 
\put(3,3){{$\lambda = 1$:}} 
\put(5.7,3){{particles, symmetries}} 

\put(7.5,1.8){\oval(6,1.5)} 
\put(3,1.6){{$\lambda = 0$:}} 
\put(4.9,1.6){{ultraparticles, ultrasymmetries}}  

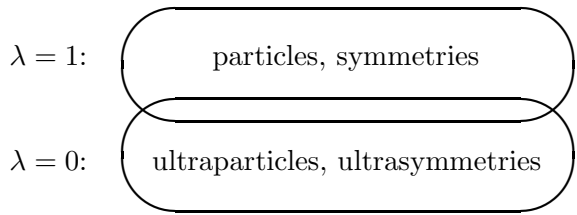
\put(2.6,0){{\small Fig.: Comparison of particle and symmetry content}}
\end{picture}}\\[2mm]
In this way, notions such as parton, colour symmetry, confinement 
etc.\ acquire an unambiguous and intrinsic meaning \cite{Bu2}. 

\section{Illustrations}
As was already mentioned, our general method has been applied to
several models \cite{BuVe2,Bu2}. A nice illustration of the scenario 
outlined in the preceding section is provided by the 
Schwinger model (massless QED in $d=2$ dimensions). It is known from 
the work of Lowenstein and Swieca that the net of observables in this 
theory is isomorphic to the net generated by a free scalar field
$\phi$ of 
mass $m > 0$. The fact that the latter net has only a single (vacuum) 
sector is then interpreted as confinement
(respectively screening) of the electric charge. It is therefore of
interest to determine the ultraparticle and ultrasymmetry content of
this model. 

To this end one has to proceed to the associated scaling algebra
$\Alu$. A typical element of this algebra is given by 
$$ \lambda \rightarrow
\underline{{A}_{ }}_\lambda = \int dx \, g(x) \, e^{\, i 
{ \int }
dy \, f(y-x) \, Z_\lambda \, \phi (\lambda y) },$$
where $f,g$ are real test functions and $Z_\lambda$ is a renormalization 
factor whose dependence on $\la$ is completely arbitrary. Any such
operator function is an element of the scaling algebra $\Alu$. For the
scaling limits of these operators in physical states one
obtains the following results, depending on the behavior
of $Z_\lambda$ for $\la \searrow 0$: 
$$ \lim_\lambda \underline{{A}_{ }}_\lambda = 
\left\{ \begin{array}{r@{\quad : \quad }l} {A_0} 
& Z_\lambda \simeq 1\\
{C} & Z_\lambda \simeq | \ln \lambda |^{- 1/2} \\
{c \, 1} & \mbox{{otherwise}} \end{array}. \right. $$  
Here $A_0$ is the operator which one obtains if one replaces 
$Z_\lambda \, \phi (\lambda y)$ in the expression for 
$\underline{{A}_{ }}_\lambda$ by the 
massless free field $\phi_0 (y)$, $C$ is some random variable which 
commutes with all operators in the scaling limit theory and $c$ is some
complex number. Hence the scaling limit theory is a central extension
of the net generated by the free massless field. 

This result shows 
that one does not need to know from the outset the (anomalous)
dimension of the field $\phi$ in order to get a well defined
limit. The theory takes care of that by itself in the sense that only
those elements of the scaling algebra have a non--trivial scaling limit where
$Z_\la$ has the appropriate asymptotic
behavior. 

In a final step one has to compute the ultrasymmetries and
ultraparticles in the scaling limit theory. It turns out that, in contrast
to the theory at scale $\la = 1$, there appear superselection sectors 
in the scaling limit which carry a charge
of electric type (in the sense that Gauss' law holds for it) and 
have a particle interpretation \cite{Bu2}. Hence in this sense the model has
a non--trivial parton structure.
 
\end{document}